\documentclass[twocolumn,prl,aps,superscriptaddress,floatfix,showpacs]{revtex4}
\usepackage{bm}
\usepackage{graphicx}

\begin{document}

\title{Phonon anomalies at the valence transition of SmS : An
inelastic
X-ray scattering study under pressure.}

\author{S. Raymond}
\affiliation{CEA-DSM/D\'epartement de Recherche Fondamentale sur
la Mati\`ere Condens\'ee / SPSMS, 38054 Grenoble, France}
\author{J.P. Rueff}
\affiliation{Laboratoire de Chimie-Physique, CNRS, 11 rue Pierre et
Marie Curie, 75231 Paris, France}
\author{M. D'Astuto}
\affiliation{European Synchrotron Radiation Facility, BP 220, 38043
Grenoble, France}
\author{D.Braithwaite}
\affiliation{CEA-DSM/D\'epartement de Recherche Fondamentale sur
la Mati\`ere Condens\'ee / SPSMS, 38054 Grenoble, France}
\author{M. Krisch}
\affiliation{European Synchrotron Radiation Facility, BP 220, 38043
Grenoble, France}
\author{J. Flouquet}
\affiliation{CEA-DSM/D\'epartement de Recherche Fondamentale sur
la Mati\`ere Condens\'ee / SPSMS, 38054 Grenoble, France}

\date{\today}

\begin{abstract}

The phonon dispersion curve of SmS under pressure was studied by inelastic x-ray 
scattering around the pressure-induced valence transition. A significant 
softening of the longitudinal acoustic modes propagating along the [111] 
direction was observed spanning a wide $q$ region from 
($\frac{2\pi}{3a},\frac{2\pi}{3a},\frac{2\pi}{3a}$) up to the zone boundary as 
SmS becomes metallic. The largest softening occurs at the zone boundary and 
stays stable up to the highest measured pressure of 80 kbar while a gradual 
hardening of the low $q$ modes simultaneously appears. This phonon 
spectrum indicates favorable conditions for the emergence of 
pressure-induced superconductivity in SmS. 

\end{abstract}


\maketitle

In the past decades, a rich variety of condensed matter phenomena (high 
temperature superconductivity, giant magnetoresistance, heavy fermion ground 
state) has been attributed to the interplay between charge, lattice and magnetic 
degrees of freedom. Pressure is a unique tool to tune the different couplings  
between these parameters. One spectacular achievement of the past years is the 
pressure-induced non-conventional superconductivity observed at the quantum 
critical point of several heavy fermion systems \cite{Mathur,Flouquet}. While 
spin fluctuations are generally believed to be responsible for the Cooper 
pairing in these compounds, superconductivity is enhanced near the charge 
instability of some of them \cite{Jaccard}. This motivated us to reinvestigate 
intermediate valence compounds where the charge fluctuations are at their 
strongest and in particular the case of SmS \cite{Wachter}. At ambient pressure, 
SmS is a semiconductor which crystallizes in the NaCl structure (black phase) 
with a divalent Sm$^{2+}$ ionic configuration (4$f^{6}$). At 6.5 kbar (at room 
temperature), it undergoes a first order isostructural phase transition 
\cite{Jama} to a metallic state (gold phase). In this phase, the Sm ion has an 
intermediate valence achieved by promoting a 4$f$ electron into the conduction 
band : Sm$^{2+}$$\leftrightarrow$ Sm$^{3+}$ + $5d$. The semiconducting state 
persists at $T$=0 up to $P_{\Delta}$=20 kbar where the sample ultimately becomes 
metallic \cite{Lapierre}. It is expected that, near $P_{\Delta}$ or at still 
higher pressure, a magnetic quantum critical point will be reached when the Sm 
ion approaches its trivalent state. The search for the related magnetic order 
and possible superconductivity in good samples is certainly an experimental 
challenge. In this paper, we focus on the lattice dynamics of SmS under pressure 
using inelastic x-ray scattering (IXS). IXS presents several advantages here 
over inelastic neutron scattering (INS), since it can be carried out on 
micron-sized samples contained in a diamond anvil cell (DAC), and does not 
require isotopic substitution. This permitted us to extend previous INS data 
obtained on the same system \cite{Mook1} at the border of the technique to much 
higher pressure (80 kbar in the present case, while the INS experiments was 
limited to 7 kbar), while keeping a good crystal quality in the whole pressure 
range. The present study focuses on the longitudinal acoustic (LA) phonon 
dispersion curve along the [111] direction. The IXS measurements show a strong 
hardening of the LA modes close to the zone center with increasing pressure 
while for the first time, significant softening is observed at the zone boundary 
and persists far above the valence transition. This provides indications of 
favorable conditions for the emergence of superconductivity in SmS at high 
pressure.

A 150$\times$100$\times$40 $\mu$m$^{3}$ platelet of SmS was cut in the 
extensively studied  batch \cite{Lapierre}
grown two 
decades ago  by F. Holtzberg using the Czochralsky method. The direction 
[110] was normal to the surface so to have the 
three principal axis in the surface plane. The sample was then loaded in a 
rhenium gasket placed in a DAC, using methanol-ethanol 1:4 as pressure 
transmitting medium. The pressure was measured on-line by conventional ruby 
fluorescence technique. The IXS measurements were carried out on the undulator 
beamline ID28 at the European Synchrotron Radiation Facility, Grenoble. The 
incident beam is monochromatized by a perfect plane Si-crystal working in 
extreme backscattering geometry at the (9,9,9) reflection (17.794 keV). The 
monochromatic beam is then focused onto the sample position by a toroidal mirror 
in a 250$\times$80 $\mu$m$^{2}$ spot. The scattered photons are analyzed by a 
bench of five spherically-bent high-resolution Si analyzers placed on a 7~m long 
horizontal arm. The analyzers are held one next to the other with a constant 
angular offset  and operate in backscattering geometry at the same reflection 
order. The energy ($\omega$) scans are performed by varying the monochromator 
temperature while keeping the analyzer crystals at fixed temperature. The 
instrumental energy resolution achieved in this configuration is 3 meV. The 
cell was mounted in a vacuum chamber positioned on the sample stage in order to 
reduce the scattering by air. Measurements were carried out at room temperature 
in transmission geometry near the $\tau$=(2,2,2) and $\tau$=(3,3,3) Bragg 
reflections. This choice is a compromise between the benefit gained by working 
at the highest momentum transfer ($\bf{Q}$) possible because of the essentially  
$Q^2$ dependence of the IXS cross section and the limited angular opening of the 
pressure cell.  Furthermore, these ($\bf{Q}$,$\omega$) regions were free from 
contamination by the diamond phonon branches. The measurement of the LA [111] 
modes was achieved by having $\bf{q}$ //  $\bf{Q}$ //  [111] where $\bf{q}$  is 
the reduced wave-vector  $\bf{q}=\bf{Q}-\bf{\tau}$, which will be expressed in 
the following as $\bf{q}$=$(q,q,q)$ with $q$ in reciprocal lattice unit 
(r.l.u.). In fact, the longitudinal condition was only fulfilled by one of the 
five analyzers, while the four others point to slightly different directions in 
$q$-space leading to a non-negligible transverse component. Transverse 
components were also observed in the pure longitudinal configuration due to the 
finite $q$-resolution and the crystal mosaicity.

\begin{figure}[ht]  
\includegraphics[width=8.5cm]{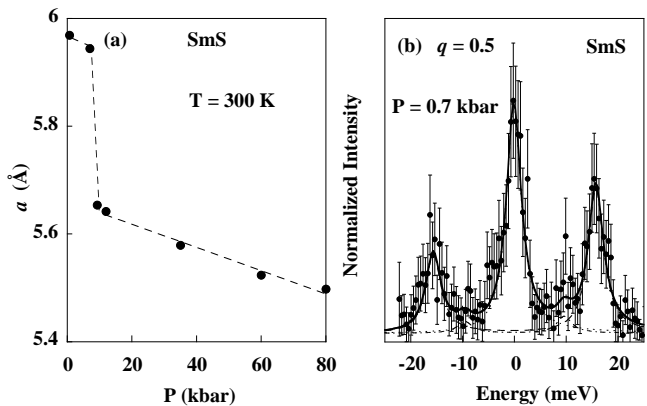} 
\caption{\label{figure1} (a) Pressure variation of the lattice parameter $a$ of 
SmS at $T$=300 K. (b) IXS spectrum measured at $\bf{q}$=(0.5,0.5,0.5) for $P$=0.7 
kbar at $T$=300 K. The lines are fit to the data including a central 
peak, a TA and a LA mode as explained in the text.} 
\end{figure}

The valence transition of SmS was monitored by the pressure-dependence of the 
lattice parameter measured on the (2,2,2) Bragg reflection as shown in 
Fig.~\ref{figure1}a. A drastic reduction of about 5 \% occurs at the valence 
transition. The overall behavior is consistent with previously reported 
measurements using x-ray diffraction in the same pressure range \cite{Keller}.
The sample mosaicity estimated by the full width at half maximum of 
the (2,2,2) Bragg reflection rocking curve increases from 0.12$^{\circ}$ at 0.7 
kbar to 0.33$^{\circ}$ at 80 kbar. This demonstrates that the crystal quality is 
still good even at high pressure, the volume collapse at the transition being 
often destructive for SmS samples. The phonon spectra were measured for each 
pressure point at 2 or 3 different crystal orientations corresponding to 10 or 
15 $q$-values for the complete set of analyzers. Typical energy scan ($\pm$25 
meV) took about 5 hours. The data were normalized to the monitor intensity. The 
high pressure spectra were further normalized to the intensity of the elastic 
peak (diffuse scattering centered at $\omega$=0) at low pressure. Both the 
Stokes and anti-Stokes parts of the spectra were fitted with Lorenztian 
weighted by the Bose thermal population factor, after deconvolution by the 
resolution function. The central contribution was also fitted by a Lorenztian. 
This is illustrated in Fig.~\ref{figure1}b at $\bf{q}$=(0.5,0.5,0.5) for $P$=0.7 
kbar at $T$=300 K. Within the accuracy of our measurement, all the phonon peaks 
were found to be resolution limited.

\begin{figure}[ht] 
\includegraphics[width=8.5cm]{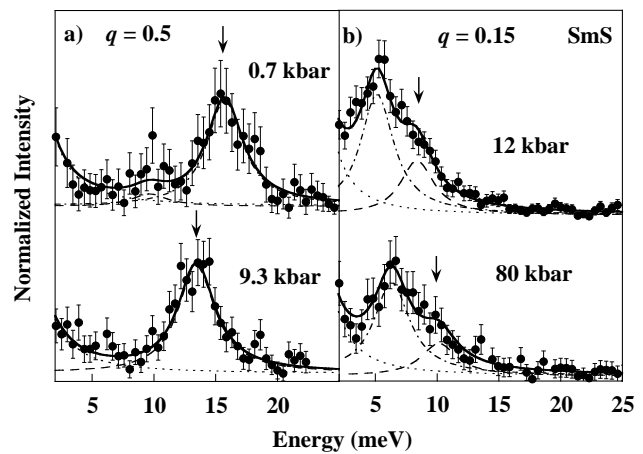}
\caption{\label{figure2} IXS phonon spectra of SmS measured (a) at 
$\bf{q}$=(0.5,0.5,0.5) before ($P$=0.7 kbar) and after ($P$=9.3 kbar) the 
valence transition at $T$=300 K. (b) at $\bf{q}$=(0.15,0.15,0.15) for $P$=12 and 
80 kbar at $T$=300 K. The lines are fit to the data as explained in the text. 
The dotted line corresponds to the central peak, the dot-dashed line to the TA 
modes and the dashed one to the LA modes. The arrows indicate the position of the 
LA[111] modes.}
\end{figure}

\begin{figure}[ht] 
\includegraphics[width=8.5cm]{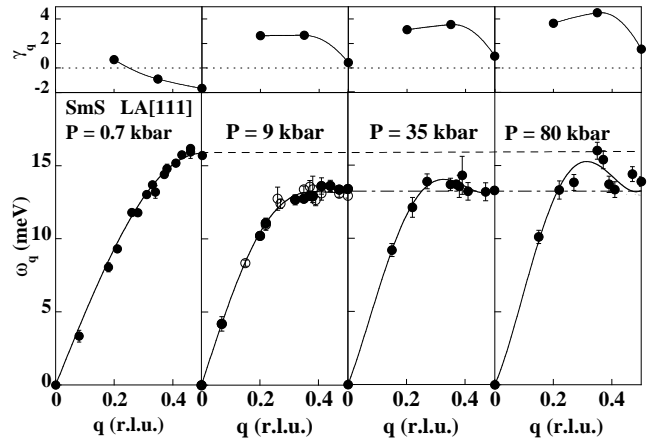}
\caption{\label{figure3} Dispersion relation $\omega_{q}$ obtained for the LA 
[111] branch of SmS at $T$=300 K for several pressures. Lines through the points 
are guides for the eyes. The open circles in the second panel correspond to data 
obtained at 12 kbar. The dashed line and the dot-dashed line shows 
the ZB value at ambient and high pressure respectively. The upper part shows 
the variation with $q$ of the mode Gr\"{u}neisen parameter $\gamma_{q}$. The 
dotted line indicates  $\gamma_{q}$=0.}
\end{figure}

The spectra measured on both sides of the valence transition at $P$=0.7 and 9.3 
kbar at the [111] zone boundary (ZB) are shown in Fig.~\ref{figure2}a. A clear 
softening from 15.7 to 13.4 meV can be observed. When pressure increases, a 
subsequent hardening occurs in the low and intermediate $q$ regions. This effect 
is illustrated in Fig.~\ref{figure2}b at $q$=0.15 r.l.u.\ for $P$=12 and 80 kbar. 
At this low $q$-value, the transverse component (dot-dashed line) is strongly 
enhanced essentially because of the mosaicity increase after the valence 
transition.
The overall dispersion curves of the LA [111] modes as a function of pressure 
obtained after fitting are summarized in Fig.~\ref{figure3}. The IXS data at 
$P$=0 are in complete agreement with the previous measurements made in the black 
phase \cite{Mook1,Birgenau} and there is in particular good reproducibility of 
the sound velocities. The sound velocity deduced from the initial slope of the 
dispersion changes from 3700 ms$^{-1}$ at $P$=0.7 kbar to 5100 ms$^{-1}$ at 
$P$=80 kbar. The softening of the phonons frequencies in the upper $q$-region 
occurs from $q$ $\approx$ 1/3 r.l.u.\ up to the ZB. At the ZB, the softening 
starts at 9 kbar and does not further evolve, once the valence transition has 
occurred. The overall behavior of the LA[111] modes is summarized in the 
pressure variation of the  mode Gr\"{u}neisen parameter $\gamma_{q}$ defined as 
$\gamma_{q}=-\partial ln\omega_{q} / \partial lnV$ ($V$ being the unit cell 
volume) and shown in the upper part of Fig.~\ref{figure3}. At low pressure, 
$\gamma_{q}$ changes sign halfway to the ZB. It becomes zero at the ZB after 
the valence transition occurs. Finally it increases for each $q$ value at higher 
pressure.

The softening of the LA [111] mode at $P>9$ kbar is found to extend over a large 
$q$-range up to the ZB. In contrast, the ZB value was not found to soften in the 
INS study limited to 7 kbar.  In order to understand this apparent discrepancy 
between the INS and IXS data, it is useful to describe the data obtained on a 
related system, Sm$_{1-x}$Y$_{x}$S, where a similar valence transition occurs 
through Y doping at about $x=x_{c}=0.15$. Among this series of compounds, 
Sm$_{0.75}$Y$_{0.25}$S in particular was extensively studied by INS 
\cite{Mook2,Mook3}, transport and thermodynamic measurements \cite{Wachter} 
since it allows one to study the  intermediate valence state by chemical 
pressure. At room temperature this compound is equivalent to SmS at $P$ 
$\approx$ 10 kbar with an intermediate valence state with 30-at.\% of Sm$^{2+}$. 
On cooling, it undergoes a valence transition at 200 K back into a state similar 
to the one of SmS at $P$=0 \cite{remarque}. The INS studies performed on this 
compound show a softening of the LA [111] modes at 300 K except at 
the ZB where a hardening is observed, similarly to the INS measurements in SmS 
at 7 kbar. Nevertheless, the ZB energy is found to decrease 
upon 
heating from 100 K to 300 K similarly to SmS under higher pressure as shown by our IXS 
data. Concerning the low $q$ modes, they have a non-monotonic behavior : they 
are soft at the transition (200 K) and stiffen at low temperature (100 K). When 
put altogether, the temperature dependence of the phonon spectrum in 
Sm$_{0.75}$Y$_{0.25}$S and the pressure dependence in SmS show distinct features 
for the mixed valence state, the nearly integer valence state and the transition 
regions.
\begin{table*}
\caption{Sound velocities in the black (or semiconducting) and gold (or 
metallic) phases and shifts in the phonon frequency in the [111] direction for 
Sm$_{0.75}$Y$_{0.25}$S, SmS and YS compounds. The upper and respectively lower 
part of the table is related to the low and respectively high pressure phases 
of SmS. The range of the physical parameters where the shift is observed is 
given in parenthesis. For YS, the shift between the ionic model (I.M.) 
and the experimental results (exp.) is considered.}
\begin{ruledtabular}
\begin{tabular}{lllll}
& $v_{black}$ & $v_{gold}$ & $\Delta\omega$ $q$=0.35 r.l.u. &
$\Delta\omega$
at $q$=0.5 r.l.u.\\
& (ms$^{-1}$) & (ms$^{-1}$) & (\%) & (\%)\\
\hline
SmS (INS)\cite{Mook1} & 4300 &  & -40 (0-7 kbar) & +17 (0-7 kbar)\\
SmS (INS)\cite{Birgenau} & 4000 &  &  &  \\
SmS (IXS, low $P$) & 3700 &  & -11 (0-9 kbar) & -17 (0-9 kbar)\\
Sm$_{0.75}$Y$_{0.25}$S (INS)\cite{Mook2} & 2900 (100 K) & 2600 (300
K) & -19 (100-300K) & -11 (100-300K)\\
\hline
SmS (IXS, high $P$) &  & 5100 & +14 (0-80 kbar) & -17 (0-80 kbar)\\
YS  (INS)\cite{Roed} &  & 5500 & - 25 (I.M.-exp.)& -88 (I.M.-exp.)
\label{table}
\end{tabular}
\end{ruledtabular}
\end{table*}
In order to sort out the different observed effects, we report in Table 
\ref{table} the sound velocities and the phonon frequency shifts measured in the 
[111] direction by INS on the Sm$_{1-x}$Y$_{x}$S system along with the IXS results on 
SmS.

At high pressures, our data exhibit an important hardening at low $q$ as 
expected for a normal metal. Unfortunately no ultrasonic measurements are 
available in SmS under high pressure due to the formation of microcracks in the 
samples which prevent the measurement. Detailed ultrasonic measurements are 
limited to the black phase \cite{Hailing} and show a general softening linked to 
the approach of the phase transition. Nevertheless, the large increase under 
pressure of the bulk modulus $B$ measured by x-ray diffraction up to 70 kbar is 
a good indication that phonon stiffening is expected \cite{Keller}. For a cubic 
system, the bulk modulus expresses as $B=(c_{11}+c_{12})/3$ and the sound 
velocity in the [111] direction is $v=\sqrt{(c_{11}+c_{12}+2c_{44})/3\rho}$ 
where $c_{ij}$ are the elastic constants and $\rho$ is the material density. 
Since $c_{44}$ is determined by the TA[100] modes which are not believed to be 
strongly pressure dependent \cite{Mook1,Hailing}, the bulk modulus and the [111] 
sound velocity should exhibit similar pressure dependence. As seen in Table \ref{table}, 
Sm$_{0.75}$Y$_{0.25}$S is much more compressive and not a good starting point 
for a discussion of the sound velocities in SmS under pressure.

The most important result of our study is the softening at the [111] ZB which is 
of the same order as the one observed in Sm$_{0.75}$Y$_{0.25}$S at $q$=0.5 
r.l.u.\ between 100 and 300 K. One possible interpretation of the discrepancy 
between the INS and IXS data obtained in SmS under pressure is that the INS results 
were limited to the transition or near-transition region ($P$=7 kbar) while the 
present IXS data were extended well within the metallic phase. 

On the 
theoretical side, several works aimed at calculating the phonon spectrum after 
the INS measurements performed on Sm$_{0.75}$Y$_{0.25}$S either by using 
microscopic models or more phenomelogical methods 
\cite{Entel,Benne,Biltz,Matsuura}. General arguments predict the location of the 
phonon anomaly \cite{Benne} : The valence transition induces a change of volume 
of the Sm ion that will at first affect the motion of the surrounding next 
nearest neighbor S atoms. Volume fluctuations occur in the longitudinal channel 
and in the [111] direction corresponding to the highest packing. Most of the 
models reproduce a softening halfway to the ZB since the zone boundary 
corresponds to Sm only motion. They are clearly not sufficient to explain our 
new results which show that also the ZB LA[111] mode is soft at high pressure. 
It is useful to note that the shape of the LA[111] phonon we measured at 80 kbar 
is qualitatively similar to the one of the superconducting compound YS 
\cite{Roed} (see Table \ref{table}). The softening at $q$=0.5 r.l.u.\ in this 
compound is to be related to a high density of states of $d$ electrons at the 
Fermi level \cite{Biltz}. Soft phonon modes and 
high density of states at the Fermi level favorize 
superconductivity as calculated in the strong coupling theory 
\cite{Mac}. Such scenario was already invoked for the other 
isostructural binary alloy NbC \cite{Sinha,Hanke}. 
 In Table \ref{table}, the phonon 
frequency shift for YS is given relatively to a theoretical calculation without 
electron-phonon coupling. One can formulate the hypothesis that pressure induces 
in parallel to (or in cooperation with) the change of valence of SmS an 
increased density of states at the Fermi level which produces the anomaly at 
$q$=0.5 r.l.u. This scenario reinforces the idea that the INS data obtained on 
SmS under pressure represents phenomena around the transition while the present 
IXS data characterizes the metallic phase up to high pressure and thus exhibits 
new anomalies in the phonon spectrum. 

Our IXS data obtained under pressure in SmS exhibit a softening of the LA [111] 
mode from halfway  and up to the ZB as soon as the metallic phase is reached. 
This softening, maximum at the [111] ZB is unchanged up to 80 kbar while gradual 
hardening of the low and intermediate $q$ modes occurs in parallel. This latter 
effect is to be linked to the unusual strong pressure dependence of the bulk 
modulus of SmS in the metallic phase. The former softening effects are 
attributed to the electron-phonon interaction occurring at the valence 
transition of SmS and also to an increasing density of states at the Fermi level 
at high pressure. Future studies aimed at finding pressure induced 
superconductivity in SmS are encouraged by the large phonon anomalies which 
persist up to high pressure where the Sm valence has certainly reached an 
integer value, and by the probable appearance of a magnetic phase opening the 
possibility of spin fluctuation mediated pairing.


\begin{thebibliography}{99}
\bibitem{remarque} This temperature induced valence transition is not
completely equivalent to the case of SmS under pressure since
the ground state of Sm$_{0.75}$Y$_{0.25}$S at $T$=0 is still mixed
valent (with 80-at.\% of Sm$^{2+}$).
\end{thebibliography}
\end{document}